# Use of Self-assembled Plasmonic Hole Arrays on AlGaAs/GaAs 2DEG for Large Area Terahertz Application

Che Jin Bae[1*], Gottfried Strasser[2], and Andrea G. Markelz[1]

[1]*Department of Physics, University at Buffalo, The State University of New York, Buffalo NY USA*

[2] *Institute for Solid State Electronics, Vienna University of Technology, 1040 Vienna, Austria*

*e-mail:chebae@buffalo.edu

**Abstract**

**Plasmonic detectors have the potential to provide a method of rapid spectroscopy without the need of moving mirrors or gratings. Previous measurements have demonstrated frequency tunable detection based on plasmonic excitations, however these devices were either small area, polarization dependent and/or required e-beam lithography. We demonstrate that large area high sensitivity THz plasmonic detection can be achieved using self-assembly nanosphere lithography. We achieve a submicron feature size grid covering a detector area of 4 mm$^2$. Measurements at 80 K show a large transmission change of 25% and a blue shift with decreasing aperture size due to coupling of disk lattice. The resonant frequencies of our device are function of radius, not periodicity. We also confirmed a magneto plasmon dispersion of the device. In conclusion we find that fabrication of self-assembled grids is a rapid and reliable method for plasmonic devices in the terahertz range.**

The use of tunable gated gratings on a two-dimensional electron gas (2DEG) structure is well known method for compact frequency sensitive THz detection based on the resonant absorption of the 2D plasmon [1-5]. The dependence of the resonant frequency on system size and the charge density is given by [1]:

$$f_o^2 = \frac{1}{4\pi^2}\frac{n_s e^2 q}{m^*(\epsilon_1 + \epsilon_2 \coth(qd))} \quad (1)$$

where, $n_s$ is charge density, $q = \frac{n\pi}{W}$ which $W$ is a periodicity of the grid, $\epsilon_1$ and $\epsilon_2$ are dielectric constants of the media surrounding the 2DEG, and $d$ is a separation between the gate and the 2DEG channel. As shown in Eq. (1), the resonance frequency has a strong dependence on the wavevector $q$ while the plasmon frequency is screened by the effective dielectric constant, a function of the separation $d$ and the wavevector $q$. The device using 1D metal gratings on the 2DEG substrate has been demonstrated with a resonance tunability by controlling the carrier density $n_s$ of 2DEG channel via a gate voltage applied over the entire illuminated region [2]. Even the geometry of a single metal strip offers a wavevector to operate 2D plasmon excitation in THz range [3, 4]. More recently it has been demonstrated that additional tuning can be attained by changing the device size, and therefore the wavevector $q$ in eq. (1), through depletion [5]. This kind of work has optical polarization dependence limiting light collecting efficiency due to 1D metal grating so that it requires fabricating large area 2D features to overcome the polarization dependence. In spite of tunability, the operating frequency range (400 ~ 600 GHz) of the device is still below terahertz region.

To extend the result to higher frequencies, one needs to go below micron size features of photolithography. However, overall sensitivity and imaging applications can be limited by the small detection area of these devices. In particular only a fraction of the light is detected, as the diffraction limited spot size is on the order of 2 mm. Large area detectors matched to the spot size can considerably simplify alignment and signal to noise. However to achieve submicron gratings over the order of 2 mm is a significant challenge. Large area small period gratings are possible using holographic lithography, however the polarization dependence of these detectors again can limit sensitivity [6]. For example for polarization independent thermal source, these grating based detectors will not detect half the incident radiation. Using a periodic grid grating with circular apertures would eliminate this polarization dependence. Large area devices can easily be achieved using photolithography, however typical systems are limited to 1 micron feature size. Large area devices with submicron features could be achieved using ebeam lithography, however each device would be expensive in ebeam writing time and the density of writing may be sufficient to induce damage in the 2DEG which is only 50 nm below the surface.

Here we overcome the limitations by using a self-assembly technique based on nanosphere lithography. We develop a nanosphere processing for GaAs substrates. Previously nanosphere colloidal solutions have been used to self-assemble into regular 2D and 3D structures [7-11]. Recently these structures have been used as a step in processing to form regular patterns with feature size as small as 100 nm [11]. However, large area GaAs devices have not been demonstrated previously.

The AlGaAs/GaAs heterostructures used in this work are grown by molecular beam epitaxy (MBE) at 615 °C on (100) oriented semi-insulating GaAs substrate as shown in Fig 1 (a) with the charge density density $n_s = 2.55 \times 10^{11}\ cm^{-2}$ and mobility is 180,000 $cm^2/Vs$. We use a self-assembled poly styrene sphere monolayer as a shadow mask for metallization and to make a large area regular metallic grid. The conditions for nanosphere lithography has been well established for glass and silicon substrates, however the formation of self-assembled monolayer depends critically on the hydrophilicity of the substrate, thus the growth conditions for glass will not work for GaAs. To form a large area nanosphere monolayer on the substrate by spin coating the appropriate nanosphere colloidal solution. The polystyrene sphere colloid solutions (Carboxyl Latex, d=1.4 ± 0.1 µm, 4% of weight per volume in the de-ionized water medium) is purchased from Interfacial Dynamics Corporation (Portland, OR) then further diluted in a solution of the surfactant Triton X-100/methanol (1:400 by volume) before spin-coating. The dilution factor for our process is 1:1 by volume. The surfactant assists the solutions in wetting the substrate [10]. Using a custom built spinner, we drop cast 7~9 µl polystyrene sphere solution on the 2DEG substrate. With a slow speed spinning (360 rpm ~ 420 rpm) for approximately 50 minutes, a large area self-assembled hexagonal closed pack monolayer is formed over a maximum 4 mm² area. Yields of uniform monolayers with larger than 2 mm² coverage were 80% for 420 rpm and 60% for 360 rpm. If the self-assembled monolayer is used as a metallization mask without further processing, one can achieve a Fisher pattern, that is a pattern of disconnected triangles. We require an interconnected metallization, both to achieve the necessary in-plane electric field modulation for coupling light to the 2D plasmons, and also for the depletion control of the 2DEG. In order to have an interconnected conducting sheet with circular apertures, we use Reactive Ion Etching (RIE) of the self-assembled nanosphere monolayer to reduce the sphere size and increase spacing between spheres. RIE using oxygen at low power density etches the polystyrene spheres uniformly. $O_2$ gas was used at 0.25 W/cm$^2$ [12]. In all cases, the chamber pressure was about 50 mTorr and the gas flow rate was 40 cm$^3$/min. A fully interconnected pattern can be achieved with an etching rate 50 nm/min allowing for the entire region to be used as a gate. The sample with the etched monolayer is then placed in a thermal evaporator and 50 nm of aluminum is deposited. After metallization the sample is placed in toluene and gently sonicated to lift off the nanospheres. Fig. 2(a) is a schematic of this procedure. In Fig. 2(b), we show an image of the final 2D grid metallization on the 2DEG material.

As seen the metal in Fig. 2(b) is completely interconnected. The dark gray holes are 2DEG substrate underneath the interconnected Al metallization (in white). This metallization is designed to both couple light to the 2D plasmon and as a gate to tune the charge density to tune the resonant frequency. The aperture size is 1.0 ± 0.1 µm, and the period is 1.4 ± 0.1 µm. As seen, the hole array has some point and line defects. We tested the effect of these on the resonant linewidth with THz time-domain spectroscopy (TDS). The unit cell of the structure is rhombus with the angle of each array as 60$^o$. As discussed above, the 2D hole array removes the polarization dependence of the device so that a transmission enhancement is also expected.

THz transmission measurements are performed using a standard THz-TDS system as described in Ref. 13. The sample is held either in a continuous flow cryostat, or in a magneto optical cryostat in the Faraday geometry with both the light propagation direction and magnetic field direction perpendicular to the 2DEG. The sample is mounted on a copper holder with a 2.38 mm diameter aperture. A reference substrate is mounted over an adjacent identically sized aperture. A transmission measurement consists of toggling between the reference and the sample. These grids are made on both 2DEG material and semi insulating GaAs substrates. The THz response is characterized using THz-TDS as a function of temperature and magnetic field.

Zero magnetic field THz transmission measurements demonstrate the plasmonic response for these devices. Fig. 3 (a) shows THz-TDS transmission measurements of the grid on the 2DEG material and the SI GaAs substrate. The transmission for the grid fabricated on SI GaAs substrate (shown in green) does not have any strong frequency dependent features. Whereas for the 2DEG material we find strong resonant absorption at 0.43 THz, 1.45 THz, and 2.13 THz at room temperature, indicating the 2DEG is necessary for the resonances. The absorbances strengthen and narrow with at lower temperature, consistent with broadening of plasmonic resonances due to temperature dependent scattering. In Fig. 3 (a), the THz transmission at low temperature 80 K shows the large and somewhat narrow band absorption achieved in this structure with a transmission change defined as $\Delta T/T$ is approximately 25% change in the transmission at the fundamental referencing with at 0.2 THz. There have been few reports of absolute absorption for plasmon detectors. However our 2D grid device appears to have a strong absorption in comparison with a previous work which only reported a 0.7% transmission change using a grating coupler on similar 2DEG system [14]. This large transmission change is caused by the enhanced collection efficiency due to large area of plasmonic structure and elimination of polarization dependence by fabricating 2D grids. A high frequency cut-off of linearly declined transmission line is shown through entire spectrum in Fig 3 (a). We attribute the capacitive grids effect such as a low pass filter to the polystyrene residue on the device [15]. By eliminating linear response from polystyrene residues, the normalization transmission in Fig. 3 (b) has been obtained to determine the resonant frequencies more specifically. We found a slight red shift as eliminating the residue effect so that the resonant frequency of fundamental confirmed as 0.38 THz in Fig. 3 (b).

The resonant frequency of the device based on lattice coupling to 2DEG has been demonstrated in terms of wavevector $q$ of gratings and charge density $N$ of the 2DEG. For the device with 1D metal strips having periodicity $l$ on substrate, the dispersion relation varies with device types [16]:

$$\omega_0^2 = \frac{e^2 Nq}{m^* \varepsilon_{eff}}; \quad q = \frac{n\pi}{l}, n = 1,2,3,\ldots, \qquad (2)$$

The effective dielectric constant varies according to the location of 2DEG channel. Eq. (1) can be defined when $\varepsilon_{eff} = \frac{1}{2}[\varepsilon_s + \varepsilon_b coth(qd)]$, corresponding to 1D gratings with a uniform metallization on a heterostructure 2DEG [6, 16]. We interestingly note that the resonant frequency of 1D strip grating device is mainly function of wavevector $q$, not duty cycle.

Other than the above, the geometry of the structure is the key point to determine the dispersion relation in 2D grids structure. The isolated circular 2DEG disk array has demonstrated by investigating a relation of the reduced dielectric constant considering both depolarization shift and coupling of disk lattice [17-20]. With assumption of the disk as thin oblate conducting spheroid [18, 19], the plasmon resonant frequency of 2DEG array having disk diameter $d$, periodicity $a$, and $\xi$ of the diagonal lattice tensor [20] is given by:

$$\omega_0^2 = \frac{3\pi e^2 N}{8m^* \varepsilon_{eff} d}\left(1 - \frac{2d^3}{3a^3}\xi_{x,y}\right) \quad (3)$$

In the case of non-isolated 2D circular hole array, Fetter and Heitmann have demonstrated the radius dependent dispersion relation which is given by [21, 22]:

$$\omega_0^2 = \frac{e^2 N}{2m^* \varepsilon_{eff}} \frac{i}{R}, \quad i = 1,2,3, ..., \quad (4)$$

where R is the radius of the disk and $\varepsilon_{eff}$ the effective dielectric function of the surrounding media [22, 23].

We examine this by changing the aperture size while holding the periodicity constant. Using the same processing as discussed, we vary our RIE time to achieve smaller aperture size down to 800 nm and 600 nm. As shown in Fig. 4 (a), the fundamental resonant frequencies for aperture size 800 nm and 600nm at 80 K are 0.50 THz and 0.58 THz respectively. The plasmon resonances yield 0.38 THz, 0.5 THz, and 0.58 THz for aperture size 1 μm, 0.8 μm, and 0.6 μm respectively in Fig. 4 (b). In Ref 24, a 2D grid array along with rectangular array axis has been reported with the resonant frequency form based on eq. (2) and the effective dielectric constant, defining two separations from the channel respectively to the surface ($d_1$) and to the metallization ($d_2$), $\varepsilon_{eff} = \frac{1}{2}[(\varepsilon_s + \varepsilon_b coth(qd_1)) + (\varepsilon_s + \varepsilon_b coth(qd_2))]$ [24]. The black line of the Fig. 4 (c) based on periodicity dependent plasmon dispersion relation using the form discussed above represents the same frequency along the aperture size change due to the same periodicity. Comparing both blue line (eq. (3)) and red line (eq. (4)) with the experimental data, we realized that the form of Fetter and Heitmann has better agreement with our data. Here we empirically verify that the resonant frequency of our device is inversely proportional to square root of hole radius, not periodicity.

In the presence of magnetic field, the coupling of the plasmon with the cyclotron resonance results in a splitting of the plasmon resonance as increases with magnetic field in Fig. 5(a). By applying magnetic field from 0 T to 3 T, we measured the absorption response at 10 K. And it results in a splitting of the plasmon resonance which increases with magnetic field. At the 1 T of magnetic field, the transmission around 0.5 THz split into double peaks as plasmon and cyclotron resonance. In Fig. 5(a), the plasmon resonance frequencies decreased a bit and converged at 0.3 THz as the cyclotron resonance frequencies continued to propagate right-hand side by increasing magnetic field. In early two reports, the uniformity and/or anisotropy of 2DEG array was strongly related with zero field resonant frequency varying with the applying magnetic field direction [17, 19]. In Fig. 5(b), the double peak at the zero magnetic field is in a good agreement with anisotropic 2DEG disk array aspect of the dispersion relation. The slope (black dashed line) of the dispersion relation between the applying magnetic field and the resonance frequencies of either the cyclotron or plasmon is related to the effective mass of 2DEG. The calculation of the effective mass by using slope of the dispersion relation $f_R = \frac{eB}{2\pi m^*}$ is reasonable with $0.069 m_e$.

In conclusion, we have successfully formed a self-assembled submicron hole array on the AlGaAs/GaAs 2DEG using the self-assembled nanosphere lithography. This allows us to improve detector sensitivity, and remove polarization dependence. Due to the improvement, the measurements at 80 K show a large transmission change of 25%. There exists a blue shift as decreasing aperture size. We also confirmed a cyclotron resonance dispersion relation and tunability of the device in the presence of magnetic field. The aspect of coupling of cyclotron with plasmon resonance has a good agreement with theoretical anticipation. As discussed above, our plasmonic device using the self-assembled fabrication is a rapid and reliable photo-conductive detector in the terahertz range.

We specially acknowledge Dr. Gottfried Strasser and the group members in Vienna University of Technology in Austria for providing us the high quality 2DEG material. This work has been supported by National Science Foundation (NIRT: ECS0609146).

Figure caption

Fig. 1. (a) The AlGaAs/GaAs heterostructures 2DEG used in this work is grown by molecular beam epitaxy (MBE) at 615 °C on (100) oriented semi-insulating GaAs substrate. The heterostructure was grown as Fig. 1(a): 50 nm undoped GaAs buffer layer, AlAs and GaAs smoothing superlattice, 1000 nm undoped GaAs layer, 20 nm Al$_{0.33}$Ga$_{0.67}$As spacer layer, modulation-doped Al$_{0.45}$Ga$_{0.55}$As layer with a Si-doping concentration of $1\times10^{18}$ cm$^{-3}$, and 5nm undoped GaAs capping layer. (b) Hall measurement was performed in the presence of magnetic field of 1000 G in terms of charge density (n$_s$) and mobility (µ) with illuminating light and then dark. The temperature dependent mobility and charge density measured for this material is shown in Fig. 1(b). At 80 K, the charge density is $n_s = 2.55 \times 10^{11}\ cm^{-2}$ and mobility is $180,000\ cm^2/Vs$.

Fig. 2. (a) Schematic for the fabrication and (b) microscopic images

Fig. 3. (a) Transmission measurement for bare 2DEG (green), 1 micron hole array at 293 K (red), and 80 K (blue). There's no evidence of frequency dependence at bare 2DEG transmission. The three strong resonances have been observed over 1micron hole array either room temp or low temp. (b) Normalized transmission by eliminating polystyrene residual effect.

Fig. 4. (a) Transmission for aperture size 800 nm (brown) and 600 nm (purple). (b) Fundamental frequency red shift as decreasing aperture size. (c) The calculated resonant frequency using eq. (2) with $\varepsilon_{eff} = \frac{k}{2}[(\varepsilon_s + \varepsilon_b coth(qd_1)) + (\varepsilon_s + \varepsilon_b coth(qd_2))]$ where $k$ = duty cycle is 0.55 THz (black line) for 1micron hole array. Dahl and Kotthaus form for 2DEG disk array shows blue curve but far off from the experimental data. Eq. (4) of Fetter form (red curve) has the best agreement with our data (red dots).

Fig. 5. (a) Magnetoplasmon (red arrows) coupling to the plasmon (blue arrows) is shown by applying magnetic field from 0 to 3T in step of 0.5T. (b) The slope of magnetoplasmon dispersion relation is inversely proportional to the effective mass 0.069$m_e$.

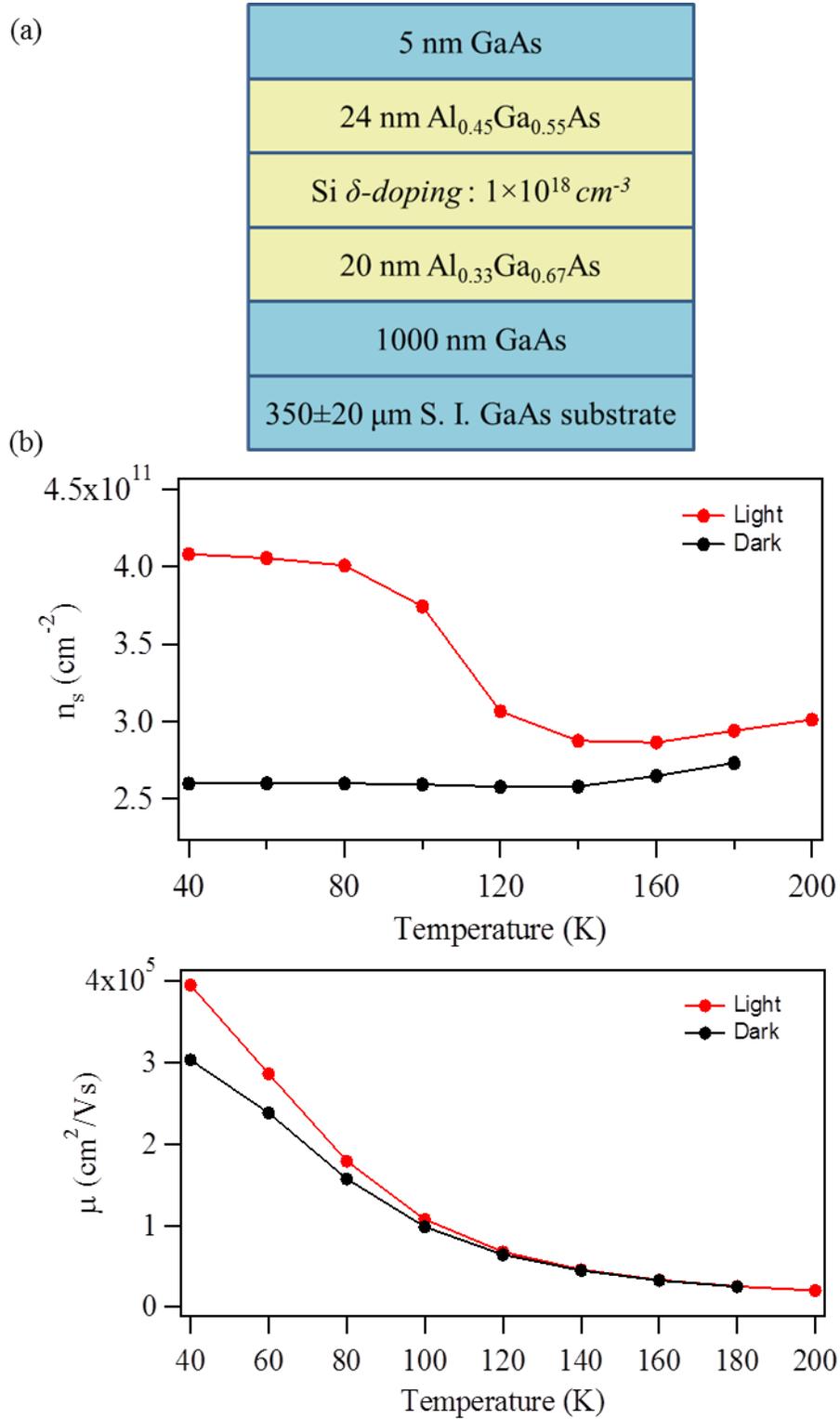

Figure 1

(a)

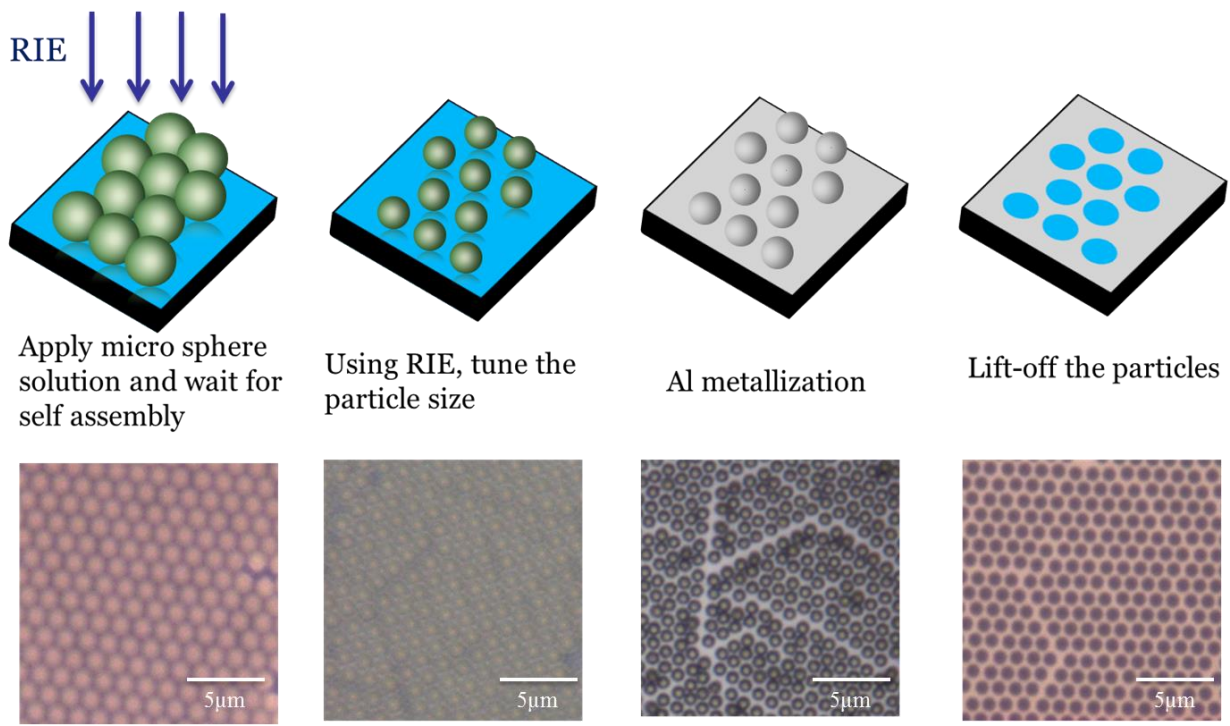

(b)

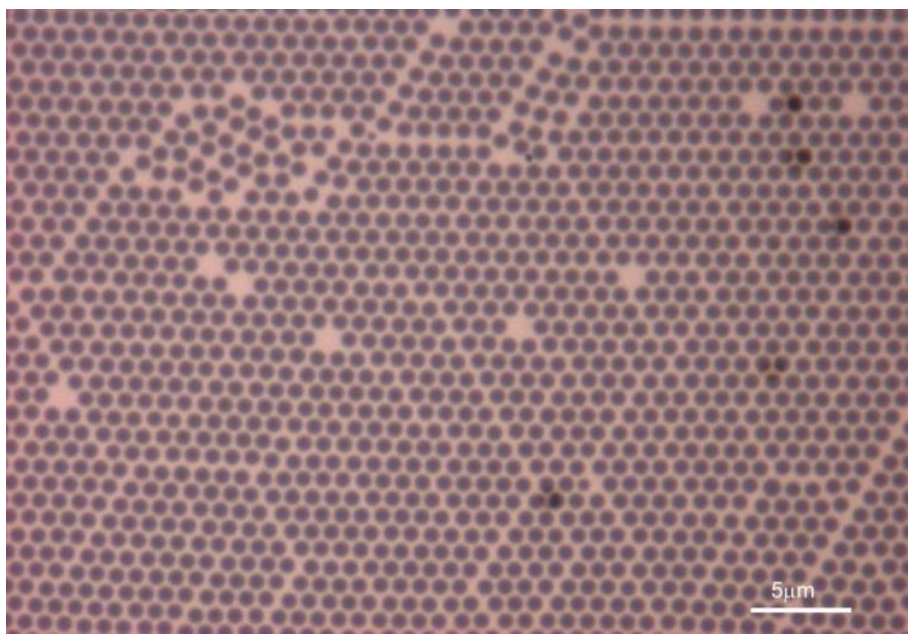

Figure 2

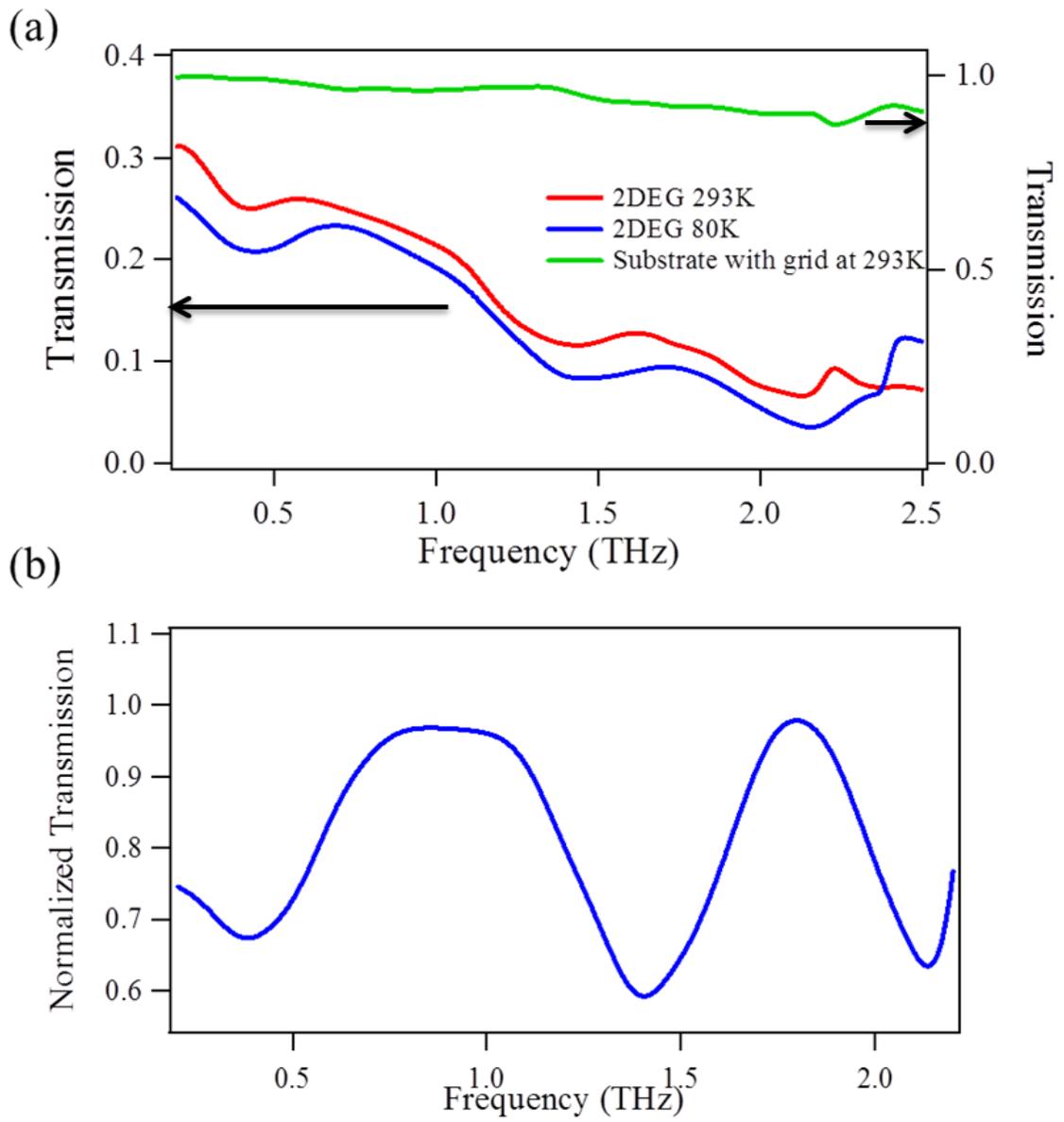

Figure 3

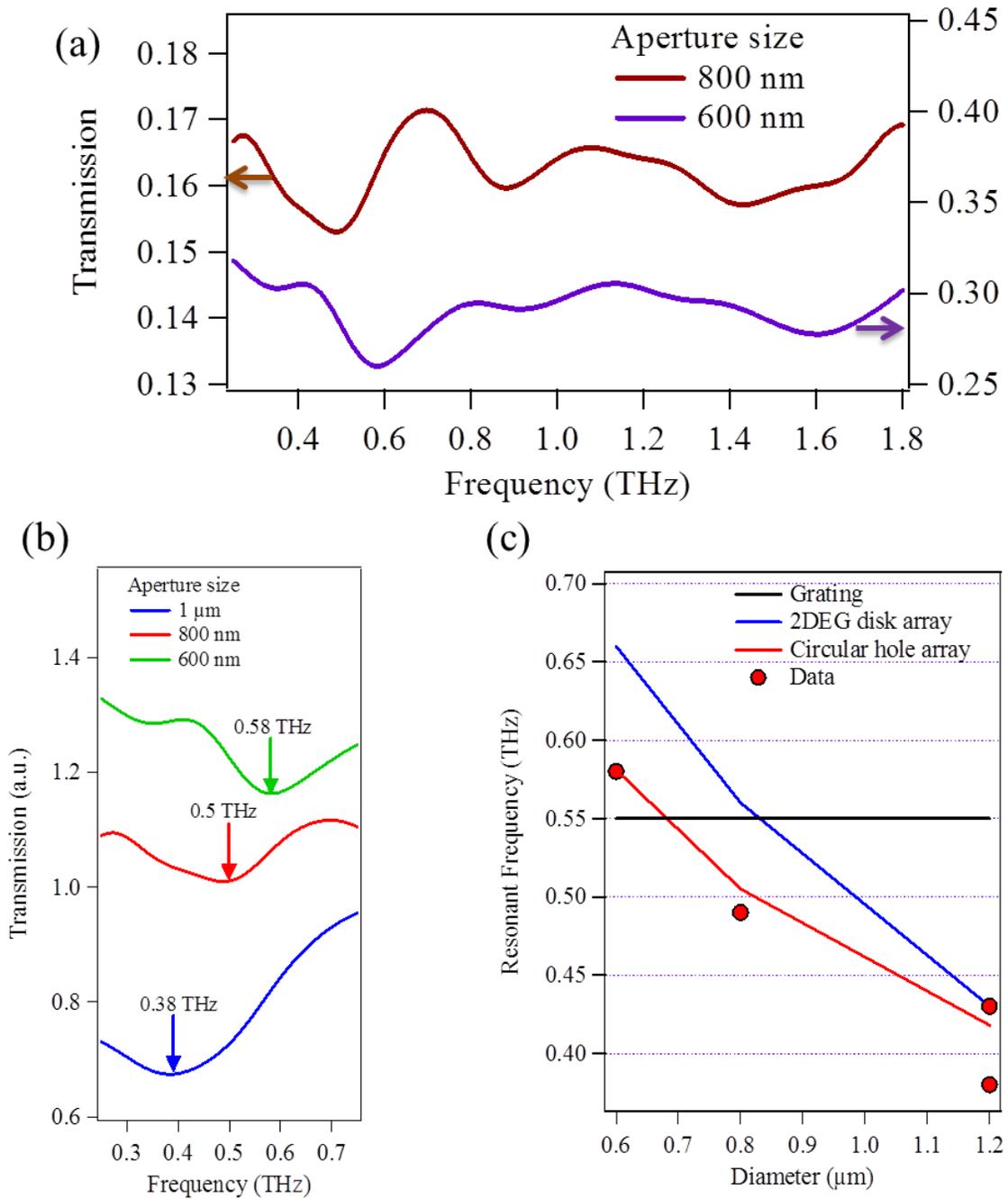

Figure 4

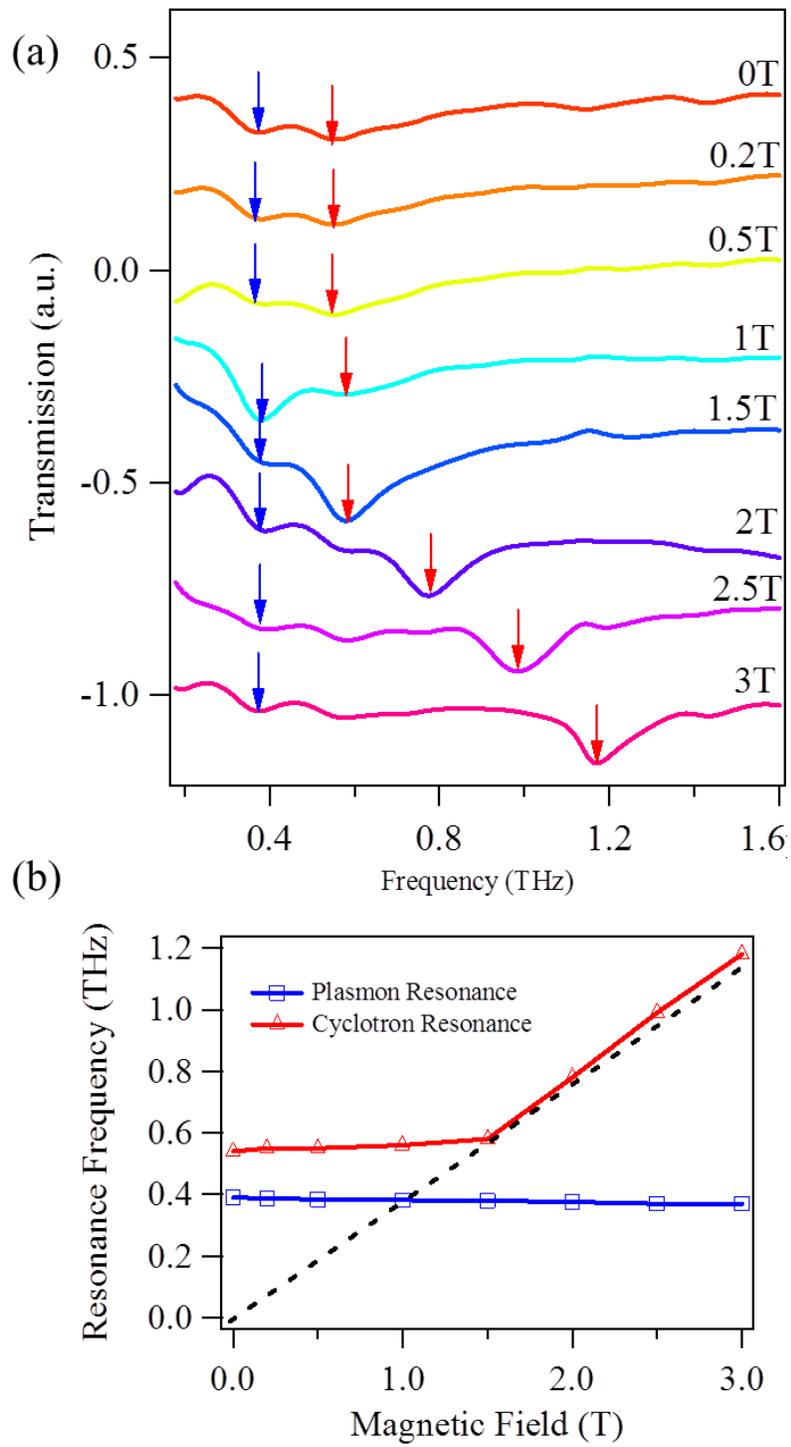

Figure 5

**Supplemental data**

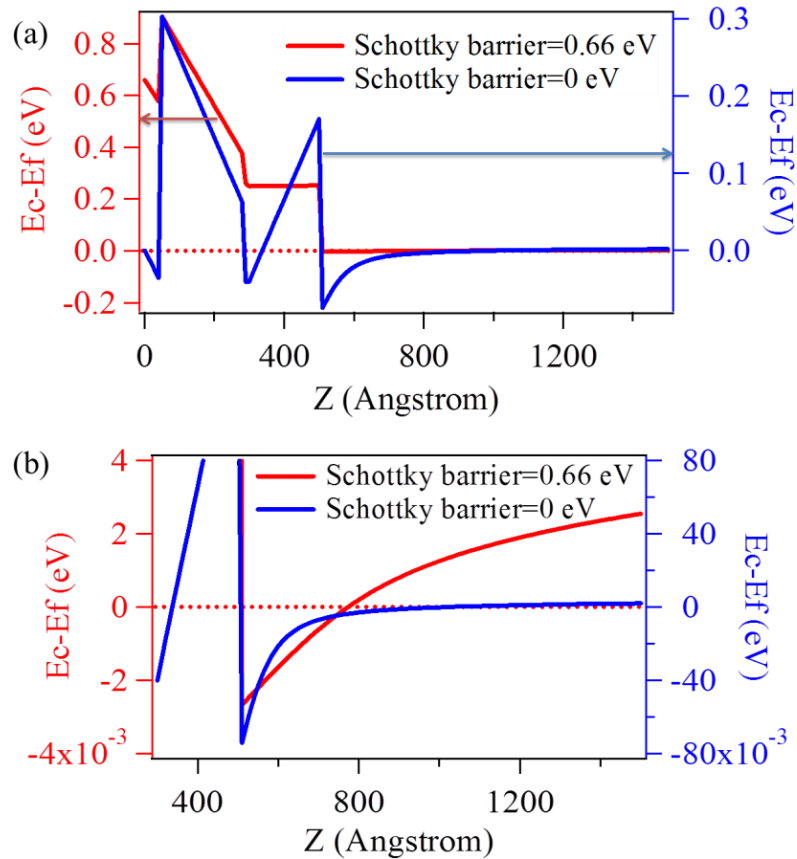

Figure 6: The band diagram of the sample structure discussed in Fig. 1 (a) is calculated by 1D Poisson equation [25, 26]. In Fig. 6 (a), the band diagram calculation without Schottky barrier (Blue line) is shifted up reasonably when considering the calculation with Schottky barrier due to Al metal contact on top (Red line). The energy level is relative value between Fermi energy (Ef) and conduction band energy (Ec) so that the zero line (Red dots) represents the Fermi energy level. Thus the band diagram demonstrates the confinement of our system. (b) To help better understanding of the confinement, we zoom in around Fermi energy level of the band diagram in Fig. 6 (a).

[25] G. L. Snider, I. –H. Tan, and E. L. Hu, J. Appl. Phys. **68**, 2849 (1990)

[26] I. –H. Tan, G. L. Snider, and E. L. Hu, J. Appl. Phys. **68**, 4071 (1990)